\documentstyle[aps,prl,multicol,epsfig]{revtex}
\begin{document}
\draft
\title{Kondo effect out of equilibrium in a mesoscopic device}
\author{S. De Franceschi$^1$, 
R. Hanson$^1$,  W. G. van der Wiel$^1$, J. M. Elzerman$^1$, J. J. Wijpkema $^1$,
 T. Fujisawa$^2$, S. Tarucha$^{2,3}$, and L. P. Kouwenhoven$^1$}
\address{{$^1$Department of Applied Physics, DIMES, and ERATO 
Mesoscopic 
Correlation Project, Delft University of Technology,
PO Box 5046, 2600 GA Delft, The Netherlands}
\newline
{$^2$NTT Basic Research Laboratories, Atsugi-shi, Kanagawa 243-0198, 
Japan}
\newline
{$^3$ERATO Mesoscopic Correlation Project, University of Tokyo, 
Bunkyo-ku, 
Tokyo 113-0033, Japan}
}
\date{\today}
\maketitle
\begin{abstract}


We study the non-equilibrium regime of the Kondo effect in a quantum dot 
laterally coupled to a narrow wire. 
We observe a split Kondo resonance when a finite bias voltage 
is imposed across the 
wire. The splitting is attributed to the creation of a double-step 
Fermi distribution function in the wire.
Kondo correlations are strongly suppressed when the voltage across the wire 
exceeds the Kondo temperature. A perpendicular magnetic field enables 
us to selectively 
control the coupling between the dot and the two Fermi seas 
in the wire. Already at fields of order 0.1 T only the Kondo 
resonance associated with 
the strongly coupled reservoir survives.   

\end{abstract}


\begin{multicols}{2}


Mesoscopic devices form a powerful tool for the study
of fundamental 
many-body phenomena. A striking example is the Kondo effect in a 
quantum-dot device, which consists of a small electronic island 
connected by tunable tunnel barriers to 
extended leads \cite{NATO}. Predicted in 1988 \cite{Theory88}, 
the Kondo effect in a quantum dot was 
observed for the first time ten years later \cite{Exp98}, leading 
to  intense research \cite{KG}.
Quantum dots offer new control to study various aspects of the 
Kondo effect.
One option is the possibility to explore non-equilibrium regimes,
e.g. by applying a 
finite voltage between the leads connected to the dot. 

At equilibrium, Kondo correlations give rise to a sharp resonance 
in the density of states (DOS). The resonance is aligned with the 
Fermi energy of the leads.
For a finite bias voltage the resonance is predicted \cite{Meir} to 
split as shown qualitatively in Fig. 1a. 
The resulting two peaks are aligned with the Fermi energies of the 
leads, and become
progressively smaller as the bias voltage is increased. 
This suppression is due to decoherence introduced by the inelastic 
scattering of electrons from the high- to the low-energy 
lead \cite{Meir,Kaminski,Rosch}.
   
Measuring differential conductance, $dI/dV$, as a function of bias 
voltage, $V$, results 
in a nonlinear characteristic with a {\it single} peak at zero 
bias \cite{Meir}. This zero-bias anomaly,
observed in a variety of experiments, is a 
characteristic signature of the Kondo effect. However, a
$dI/dV$-vs-$V$ measurement does not provide enough information to 
extract the $V$-dependent DOS.
In particular, whether the Kondo resonance is indeed split and how the 
resulting double-peak DOS evolves with bias is an issue that has 
not yet been experimentally 
investigated. Whether Kondo correlations survive at voltages larger 
than the Kondo 
temperature at equilibrium, is a question that has recently raised 
considerable debate \cite{Kaminski,Rosch,Coleman,Wen,Lee}.  
\begin{figure}[b]
\begin{center}
\centerline{\epsfig{file=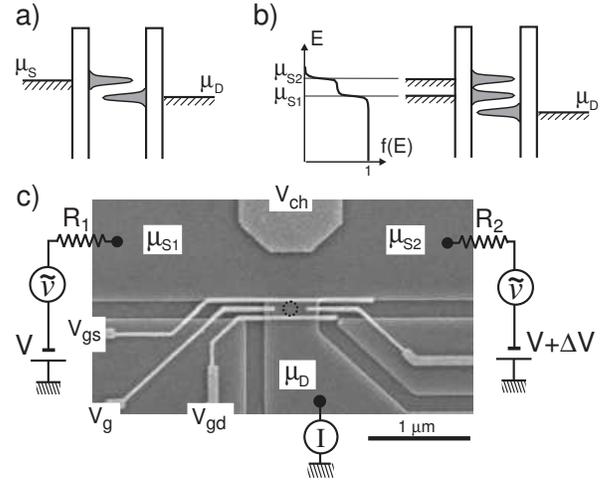, width=8cm}}
\caption{(a) Splitting of the Kondo resonance due to finite bias voltage 
across the quantum dot. The resulting split peaks in the density of states
are aligned with the Fermi energies of the leads. (b) Additional splitting 
due to a double-step distribution function in the source reservoir. 
(c) Scanning electron micrograph of the device and measurement scheme. 
Light grey corresponds to metallic gates, dark grey to etched regions.}
\end{center}
\end{figure}
Some authors \cite{Coleman,Wen} have argued the possibility of 
a two-channel Kondo effect that would develop at finite bias. 
A direct measurement of the DOS would contribute important information to 
understand Kondo physics out of equilibrium. 
Recent theoretical papers \cite{Sun,Lebanon} have proposed to use 
three-terminal quantum dots in which one of the leads serves as a 
weakly coupled electrode to probe the local DOS. 
Here we address this issue from a different angle.\\   
\indent
We couple a quantum dot to the middle of a quasi-ballistic quantum wire 
that is brought out of equilibrium by an applied bias voltage. 
The finite voltage creates 
inside the wire two ensembles of electrons (i.e. right and left movers) 
whose quasi-electrochemical potentials, $\mu_{S1}$ and $\mu_{S2}$, 
are defined by the electron reservoirs connected to the wire. 
Accordingly, the electron distribution function in the wire 
develops two steps 
located at $\mu_{S1}$ and $\mu_{S2}$ \cite{Pothier,dePicciotto} 
(see Fig. 1b, left side).
Our main goal is to show that the Kondo resonance splits 
for $\mu_{S1} \neq \mu_{S2}$ (see Fig 1b, right side).
\begin{figure}[b]
\begin{center}
\centerline{\epsfig{file=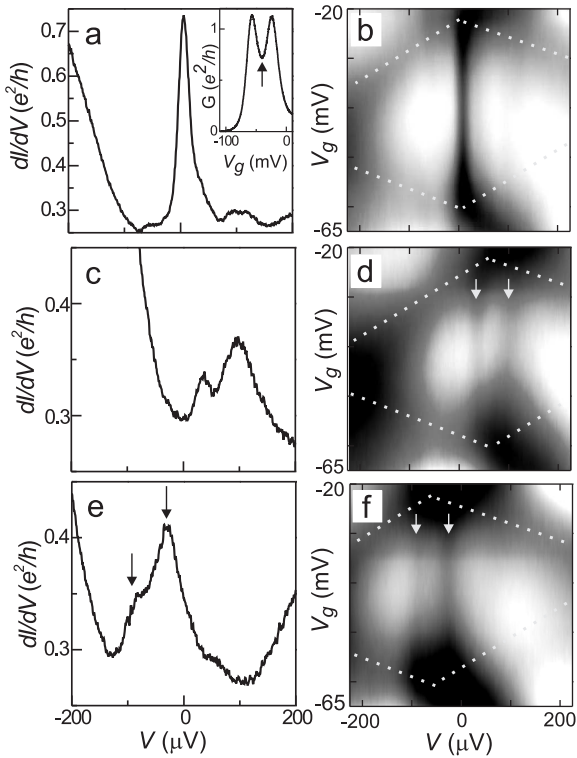, width=7.8cm, clip=true}}
\caption{(a,c,e) Differential conductance, $dI/dV$, vs bias voltage, $V$, 
across the dot. The plunger gate is at $V_g = -40$ mV. The voltage $\Delta V$ 
between source 1 and 2 is set to 0 (a), $-$150 (c), and 150 $\mu V$ (e). 
Inset to (a): linear conductance, $G$, vs $V_g$ around the Kondo valley 
(see arrow).
(b,d,f) $dI/dV$ vs ($V$,$V_g$) on gray scale. Dark gray corresponds to 
large $dI/dV$. $\Delta V = 0$ (b), $-$150 (d), and 150 $\mu V$ (f). Dotted 
lines identify the edges of the Coulomb diamond. The Kondo anomaly shows up as 
a single vertical line at zero bias in (a), and as a vertical 
split line (see arrows) in (d) and (f).  
}
\end{center}
\end{figure}
 

The sample designed for this experiment was fabricated on a GaAs/AlGaAs 
heterostructure containing a two-dimensional electron gas (2DEG) 
90 nm below the surface. (The experiment was reproduced on a sample 
with similar geometry, fabricated on a different heterostructure.)
As shown in Fig. 1c, the device has a small quantum dot 
(indicated by a dotted circle), formed inside a 500-nm-wide 
region defined by shallow etching. Confinement in the longitudinal 
direction is accomplished by applying negative gate voltages, 
$V_{gs}$ and $V_{gd}$. 
These gate voltages define the tunnel coupling between the quantum 
dot and its 2DEG leads. A 
third gate voltage, $V_g$, is used to control the electrostatic 
potential on the dot. 
On the top side, 
the dot is connected to the middle of a short quantum wire formed 
in the 2DEG 
by applying a voltage $V_{ch} = -0.4$ V to a large gate coming from 
above. The wire then has a resistance, $R_{ch}$, of 
about 1.3 k$\Omega$, corresponding to $\sim$10 spin-degenerate 
propagating modes.
There are in total three regions of 2DEG connected by ohmic contacts 
to the external circuitry. 
The bottom 2DEG (drain) is connected to ground via a current-voltage 
converter that measures the current through the quantum dot.   
The other two regions, source 1 and source 2, are connected to 
dc voltages $-$$V$ and $-$($V+\Delta V$). 
 
An additional ac voltage $\tilde{v}$ with a rms amplitude 
of 2.5 $\mu$V is applied to both source
leads allowing for lock-in measurements.
The resistances of the leads, $R_1$ and $R_2$, are in 
the 0.5$-$0.8 k$\Omega$ range, including the 
contribution from the respective 2DEGs.
We call $\mu_{S1}$ and $\mu_{S2}$ the Fermi energies of the 
source 2DEGs, and $\mu_D$ the Fermi energy of the drain 2DEG.\\
\indent
Measurements were done in a dilution refrigerator at 15 mK.
The quantum dot parameters ($V_{gs}$, $V_{gd}$, and $V_g$) were 
tuned to achieve a robust 
Kondo regime. In the inset to Fig. 2a we show the linear 
conductance, $G$, of the dot as a function of $V_g$ ($V$  
and $\Delta V$  are set to zero).  
The Kondo regime (see arrow) takes place between the two Coulomb 
blockade peaks. 
Here the valley conductance is enhanced.
By sweeping $V$ for $\Delta V = 0$ we measure  $dI/dV$ as a function of 
bias voltage $V$ across the dot. 
As in several earlier experiments, we find a narrow
peak at zero bias (Fig. 2a) which reflects the existence of a 
Kondo resonance at the Fermi energy.  
The peak has a full width at half maximum, $w_0 \simeq 25$ $\mu$V, 
yielding a Kondo temperature $T_K \sim e w_0/k_B = 0.3$ K.\\
\indent
In Figs. 2b, 2d and 2f we show on grey scale $dI/dV$ vs ($V$,$V_g$). 
Each plot is obtained from many $dI/dV$ vs $V$ traces taken for a set of 
closely spaced $V_g$ values 
ranging between -65 and -20 mV. Figure 2b, taken at $\Delta V =0$, 
shows a Coulomb diamond (indicated by dotted lines) with a clearly 
distinguishable peak at zero bias.
A non-zero $\Delta V$ results in a current $I_{12}$ between 
source 1 and 2, and a finite voltage $(\mu_{S2} - \mu_{S1})/e$ 
across the wire.
Hence the quantum dot faces, at its topside, a non-equilibrium 
electron distribution with two 
quasi-Fermi energies, $\mu_{S1}$ and $\mu_{S2}$, associated with 
the carriers from the 
left and right reservoir (see Figs. 1b,c).
Figure 2c shows a $dI/dV$ vs $V$ trace taken in the middle of the 
Kondo valley for $\Delta V=-150$ $\mu$V. 
Strikingly, the Kondo peak has split due to the applied 
voltage $\Delta V$.  
The splitting is seen all over the Coulomb diamond 
as shown in Fig. 2d. 
A similar result is found for $\Delta V = 150$ $\mu$V (Figs. 2e,f). 
In this case, however, the split peaks are not 
fully resolved \cite{note1}.
As we show below, the split peaks occur at those voltages, $V$, 
for which $\mu_{S1}$ or $\mu_{S2}$ line up with $\mu_D$. 
This represents the first 
observation of a bias-induced splitting of the Kondo resonance.\\
\indent
In Fig. 3a we plot several $dI/dV$ vs $V$ traces taken again 
in the middle of the Kondo valley, but now for different $\Delta V$ 
between -300 and 300 $\mu$V.
To compensate for the series resistance $R_1$, each trace has been 
shifted horizonally by as much as 
$R_1 I_{12}=R_1 \Delta V/(R_1+R_2+R_{ch})$, 
i.e. $V \rightarrow V + 0.20 \Delta V$. Then $V$ becomes the 
voltage $\mu_{S1}/e \simeq (\mu_{S1} - \mu_{D})/e$ (because of the 
relatively small current through the drain lead we can neglect 
the voltage drop on its series resistance and assume $\mu_D=0$).

\begin{figure}[t]
\begin{center}
\centerline{\epsfig{file=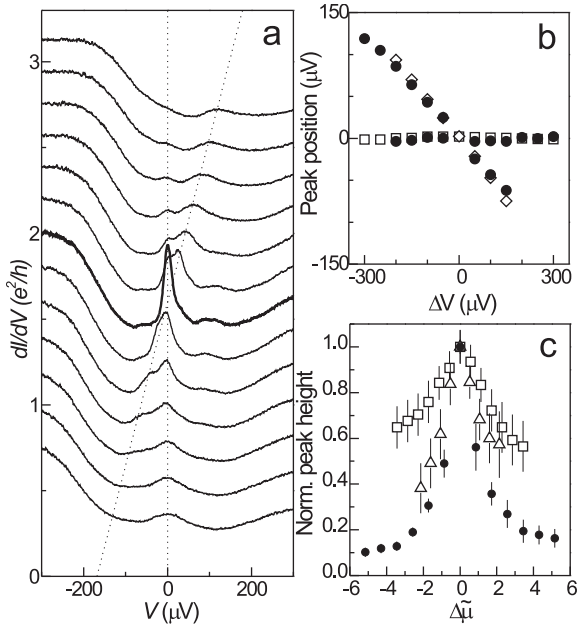, width=8cm, clip=true}}
\caption{
(a) $dI/dV$ vs $V$, for $\Delta V$ between -300 $\mu$V (top trace) 
and 300  $\mu$V (bottom trace) in steps of 50 $\mu$V. Curves are 
vertically offset by $0.2 e^2/h$. The voltage scale is shifted 
by $0.20 \Delta V$, to compensate for
the voltage drop on $R_1$. 
The zero-bias Kondo peak at $\Delta V = 0$ (thick trace) splits and weakens
at finite $\Delta V$. Dotted lines are drawn to emphasize the splitting. 
(b) $V$-position of the Kondo peaks at $B=0$ (solid circles), 
$B=-0.2$ T (open squares), and $B=0.2$ T (open diamonds). 
(c) Height of the Kondo peaks, normalized to the $\Delta V=0$ value, 
vs $\Delta \tilde{\mu} \equiv (\mu_{S2} - \mu_{S1})/e w_0$. $w_0$ is the full 
width at half maximum of the Kondo peak at $\Delta V = 0$.  For $B=0$ 
(solid circles), peak heights are obtained from fitting to a double 
Lorenztian and a linear background. 
For each $\Delta V$ we plot only the height of the largest peak. 
For $B=-0.2$ (open squares) 0.2 T (open triangles), peak heights are obtained 
from fitting to a Lorentzian and a 5th order polynomial background.      
}
\end{center}
\end{figure}

The splitting of the Kondo resonance is proportional to $\Delta V$ 
as emphasized by the dotted lines in Fig. 3a. One of the split 
peaks stays at zero bias, corresponding to the lineup condition 
between $\mu_{S1}$ and $\mu_D$. The other peak moves with $\Delta V$ 
following the condition 
$V=  (\mu_{S1} - \mu_{S2})/e = R_{ch} \Delta V/(R_1+R_2+R_{ch})$. 
This corresponds to the alignment of $\mu_{S2}$ with $\mu_D$. \\
\indent
Increasing $| \Delta V |$ (i.e. moving to the top or bottom 
of Fig. 3a) results in a suppression of the split Kondo resonance.  
The suppression is due to inelastic scattering processes in the wire 
which transfer electrons from the high- to the low-energy Fermi reservoir.
On what voltage scale is the Kondo resonance washed out?  
To address this question we plot normalized peak height 
as a function of $\Delta \tilde{\mu} \equiv (\mu_{S2}-\mu_{S1}) / w_0$ 
(see solid circles in Fig. 3c). Since $e w_0 \sim k_B T_K$, 
we are scaling the voltage across the wire in units of the 
Kondo temperature. 
For $| \Delta \tilde{\mu} | \simeq 1$ 
(i.e. $| \mu_{S2}-\mu_{S1}| \sim k_B T_K$) the peak height 
becomes half of the value for  
$\Delta V = 0$, which means that the Kondo resonance is suppressed on 
a voltage scale $k_B T_K/e$. 
For $|\Delta \tilde{\mu}| > 2$ we have fitted the data to the 
analytical function 
$A \, \mbox{ln}^{-2}(\alpha |\Delta \tilde{\mu}|) (1+2 \, 
\mbox{ln}^{-1}(\alpha |\Delta \tilde{\mu}|))$, 
derived in Ref. \cite{Rosch} in the limit $|\Delta \tilde{\mu}| \gg 1$. 
Although the latter condition is not fully satisfied we find good agreement 
for $A \sim 1$ and $\alpha \sim 5$.\\ 
\indent
We now discuss the effect of a magnetic field, $B$, perpendicular to the 2DEG.
Already for fields of order 0.1 T, the Lorentz force causes a significant 
shift in the electron waves traveling through the wire. (We note that the 
magnetic length for $B=0.1$ T is about 80 nm, i.e. a few times 
smaller than the effective width of the wire.) For $B < 0$, 
left moving electrons are pushed away from the dot, whereas right movers 
travel closer to it (Fig. 4a). The situation is reversed for $B>0$ (Fig. 4b).\\
\indent
In Fig. 4c and 4d we show $dI/dV$ vs $V$ traces taken for $B=-0.2$ 
and 0.2 T, respectively. As in Fig. 3a, $V_g$ is adjusted to the middle 
of the Kondo valley, $\Delta V$ is varied between -300 and 300 $\mu$V. 
The voltage is compensated by $0.20 \Delta V$ for the series resistance $R_1$.
As opposed to the results for $B=0$, no splitting is observed 
at finite $\Delta V$. We explain such a different behavior as follows. 
At $B= -0.2$ T only right movers
couple effectively to the dot, while the coupling for left
movers is strongly suppressed (see Fig. 4a). As a result, most of the 
current through the dot comes from the strongly coupled reservoir 
(see inset to Fig. 3c and
relative caption) and only the Kondo peak corresponding to 
the lineup between $\mu_{S1}$ and $\mu_D$ survives. 
Indeed the Kondo peak stays at $V=0$, irrespective of $\Delta V$. 
For $B=0.2$ T, only electrons from the right reservoir couple 
effectively to the dot giving rise to Kondo effect. 
The resulting Kondo peak occurs when $\mu_{S2}$ is aligned 
with $\mu_{D}$, and its position shifts proportionally to $\Delta V$.\\
\indent
In the inset to Fig. 4d we show that the coupling of the weakly coupled 
lead can be restored by making the voltage $V_{ch}$ more negative. 
This reduces the effective width of the wire pushing together left 
and right movers. 
The splitting of the Kondo resonance by a finite $\Delta V$ is recovered 
for $B=0.2$ T, similarly to the case $B=0$.\\
\indent
Finally, we compare quantitatively results for $B=\pm 0.2$ T and $B=0$.
Figure 3b shows the $V$-positions of the Kondo peaks as obtained from 
Figs. 3a, 4c, and 4d. Split peaks from Fig. 3a (solid circles) match 
well with the 
single peaks from Figs. 4c and 4d (open squared and diamonds). 
This corroborates the interpretation given earlier that split peaks occur 
when $\mu_D$ lines up with $\mu_{S1}$ and $\mu_{S2}$.

\begin{figure}[t]
\begin{center}
\centerline{\epsfig{file=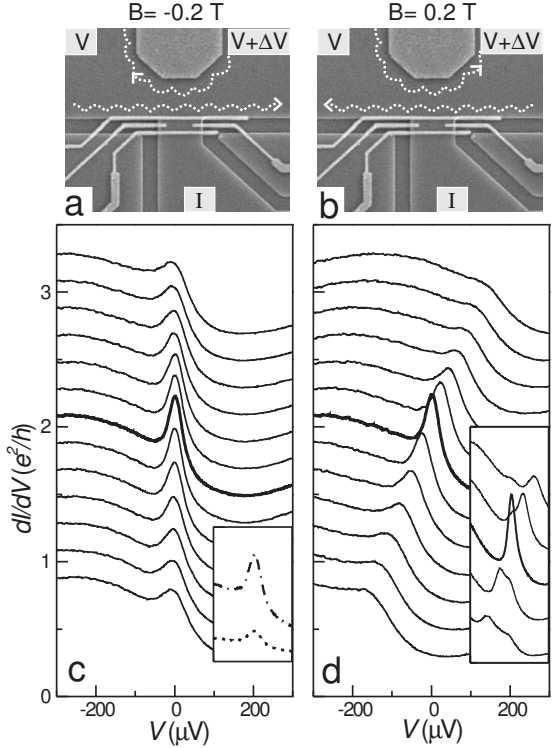, width=7.5cm, clip=true}}
\caption{
(a) White dotted lines depict electron motion through the wire 
in presence of a perpendicular magnetic field $B=-0.2$ T. The Lorentz 
force pushes right movers down, closer to the dot, and left movers up, 
away from it.
An opposite scenario occurs for $B=0.2$ T as shown in (b). 
(c,d) $dI/dV$ vs $V$, 
for $\Delta V$ between -300 $\mu$V (top trace) and 300  $\mu$V 
(bottom trace) in steps of 50 $\mu$V. 
Curves are offset vertically by $0.2 e^2/h$ and 
horizontally by $0.20 \Delta V$. 
Inset to (c) shows that most of the current through the dot comes from 
the strongly coupled lead (i.e. source 1 for $B=-0.2$ T). The dot-dashed 
(dotted) trace is $dI/dV$ vs $V$, measured by lock-in technique with the 
ac voltage excitation, $\tilde{v}$, applied only to the left (right) reservoir. 
This allows us to directly measure the contribution from the left (right) 
reservoir. The horizontal axis runs from -100 to 100 $\mu$V, the vertical 
axis from 0 to $e^2/h$. Inset to (d) shows $dI/dV$ vs $V$ for $B=0.2$ T,
$\Delta V$ from $-$150 (top) to 150 $\mu$V (bottom) in steps of 50 $\mu$V, 
and a more negative voltage $V_{ch}=-1.5$ V on the top gate (see Fig. 1c). 
Horizontal scale: from $-150$ to 150 $\mu$V. 
Vertical scale: from 0 to 2 $e^2/h$. Traces are vertically offset by 0.3 $e^2/h$. 
}
\end{center}
\end{figure}

In Fig. 3c we compare the $\Delta V$-dependence of the peak height at 
zero and finite $B$. The suppression of the Kondo resonance becomes 
considerably less severe at finite $B$. This increased robustness stems 
from the displacement between left and right movers which results in a 
lower rate for inelastic scattering events. Negative magnetic fields 
appear to be more effective in this respect \cite{note1}.\\ 
\indent
In conclusion, we have shown that a voltage difference  
between two Fermi seas coupled to a quantum dot causes  
the Kondo resonance to split, in agreement with theory. 
Split peaks are strongly suppressed for voltage differences  
much larger than 
$k_B T_K/e$. A perpendicular magnetic field has proved 
a useful tool to control the 
relative coupling of the leads. 
This provides a way to obtain a three-terminal quantum dot in which one of 
the leads is weakly coupled and can be used as a probe for the local DOS
\cite{Sun,Lebanon}. 
\\ 
\indent
We thank Yu. V. Nazarov, J. Kroha, Y. Meir, L. I. Glazman, J. E. Mooij,
T. Hayashi, B. van der Enden, and R. Schouten for discussions and help.
This work was supported by the Specially Promoted
Research, Grant-in-Aid for Scientific Research, from the Ministry of
Education, Culture, Sports, Science and Technology in Japan,
the Dutch Organisation for
Fundamental Research on Matter (FOM), the DARPA-QUIST program 
(DAAD19-01-1-0659), and the EU via a TMR network (ERBFMRX CT98-0180).


\end{multicols}


\begin{references}
\bibitem{note1} This discrepancy may be attributed to the lack of 
left-right symmetry in the device.
\end{references}
\end{document}